\documentclass[twocolumn,showpacs,preprintnumbers,amsmath,amssymb,superscriptaddress,floatfix]{revtex4}
\usepackage{graphicx}
\usepackage{amsmath}
\usepackage{bm}
\usepackage{mathrsfs}
\usepackage{color}
\usepackage{slashed}
\usepackage{dcolumn}

\newcommand{\bald}[1]{{\bf #1}}

\newcommand{\cur}[1]{\mathscr{ #1}}
\newcommand{\eqf}[1]{\begin{equation}\begin{split}#1\end{split}\end{equation}}

\begin{document}

\title{Thermal field theory derivation of the source term induced by a fast parton from the quark energy-momentum tensor}

\author{R. B. Neufeld \\
{\it Los Alamos National Laboratory, Theoretical Division, MS B238, Los Alamos, NM 87545, U.S.A.}}

\date{\today}

\begin{abstract}
I derive the distribution of energy and momentum transmitted from a fast parton to a medium of thermalized quarks, or the source term, in perturbative thermal field theory directly from the quark energy-momentum tensor.  The fast parton is coupled to the medium by adding an interaction term to the Lagrangian.  The thermal expectation value of the energy-momentum tensor source term is then evaluated using standard Feynman rules at finite temperature.  It is found that local excitations, which are important for exciting an observable Mach cone structure, fall sharply as a function of the energy of the fast parton.   This may have implications for the trigger $p_T$ dependence of measurements of azimuthal dihadron particle correlations in heavy-ion collisions.  In particular, a conical emission pattern would be less likely to be observed for increasing trigger $p_T$.  I show that the results presented in this paper can be generalized to more realistic modeling of fast parton propagation, such as through a time dependent interaction term, in future studies.
\end{abstract}


\maketitle

At sufficiently high temperature, quantum-chromodynamic (QCD) matter is expected to undergo a transition from colorless hadrons to a state of deconfined quarks and gluons known as the {\it quark-gluon plasma} (QGP) \cite{Shuryak:1980tp}.  Experimental results from the Relativistic Heavy-Ion Collider (RHIC) indicate that the QGP has been formed in relativistic heavy-ion collisions \cite{Arsene:2004fa}.  The results suggest that the QGP formed at RHIC center-of-mass energies may behave as a nearly ideal fluid \cite{Romatschke:2007mq}.  Another striking result \cite{Adcox:2001jp} is that highly energetic, or {\it fast}, partons appear to lose a significant amount of energy to the medium in a process known as {\it jet quenching} \cite{jet1}.  In the light of these observations, an interesting problem is to calculate how the QGP responds to a propagating fast parton.

This problem has gained attention due to experimental measurements of azimuthal particle correlations associated with high $p_T$ triggers in heavy-ion collisions that display a double-peaked or conical structure \cite{machexp}.  These measurements may reflect the interaction of the jet with the medium and proposed explanations for the structure have included large angle gluon radiation \cite{angle}, Cerenkov radiation, \cite{cerenk}, colored wakes \cite{wake}, and perhaps most commonly, Mach cone shockwaves excited in the bulk medium by fast partons \cite{machprop1,machprop2}. Other explanations, which do not reflect the interaction of the jet with the medium, such as fluctuating initial conditions and triangular flow have also been proposed \cite{Alver:2010gr}.  Whatever the nature of the conical structure associated with high $p_T$ triggers at RHIC is, the QGP response to a fast parton remains an important topic, particularly in the light of the surplus of high transverse momentum probes available to the newly online heavy-ion program at the Large Hadron Collider as compared to RHIC.  For this reason, the experimental handle on medium response to fast partons should only improve.

Theoretical investigation of the QGP response to fast partons has mostly been phenomenological, with many studies using hydrodynamics to model the medium response \cite{machphenom}.  Even within the framework of hydrodynamics one still must specify how the fast parton excites the medium, that is, the {\it source term}.  The source term, here denoted $J^\nu$, couples to the energy-momentum tensor (EMT, denoted $T^{\mu\nu}$ in what follows) as $\partial_\mu T^{\mu\nu} = J^\nu$ and describes the flow of energy and momentum density between an external source and the medium.  Previous studies have found that the azimuthal dihadron correlation spectrum associated with a fast parton depends sensitively on the form of the source term \cite{machprop1,neufrenk}.  It has been shown that diffusive momentum, which flows in the direction of the source parton's velocity, tends to fill up any double-peaked conical structure in the final spectrum \cite{Betz:2008wy,Betz:2008ka}.  These observations suggest that the shape of azimuthal dihadron correlations may be a powerful probe of the microscopic interaction of energetic partons with the QGP.  There has been little rigorous theoretical investigation of the source term generated by a fast parton in QCD.  However, a calculation of the source term generated by a fast parton for a perturbative QGP within kinetic theory was performed in \cite{kineticsource}.

Ideally one would like to calculate the QGP response to a fast parton from first principles, such as in the investigation of heavy quark propagation through a strongly coupled thermal plasma done by Friess {\it et. al} using the AdS/CFT correspondence \cite{adscft}.  The authors evaluated the stress-tensor within the context of linearized gravity.  It was later shown that the solution matched well with a nearly ideal linearized hydrodynamic calculation up to distances of about $2/T$ away from the heavy quark \cite{adshyd}.

\begin{figure*}
\centerline{
\includegraphics[width = 0.25\linewidth]{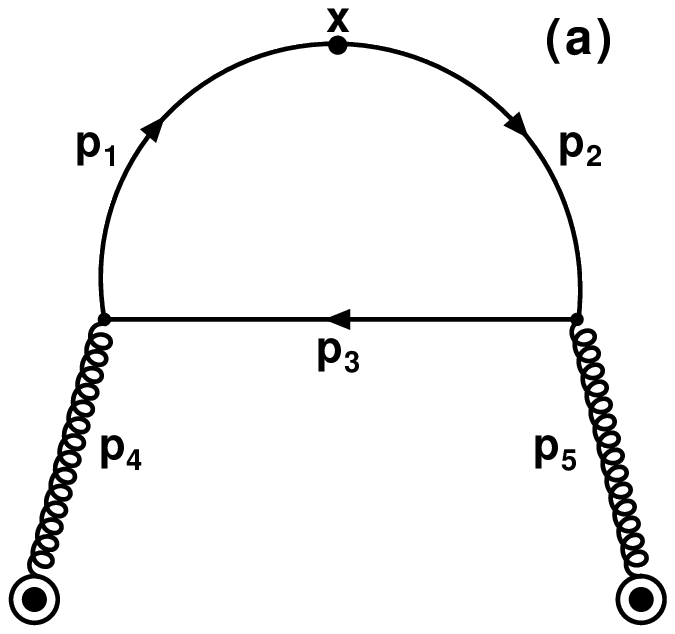}\hskip0.15\linewidth
\includegraphics[width = 0.26\linewidth]{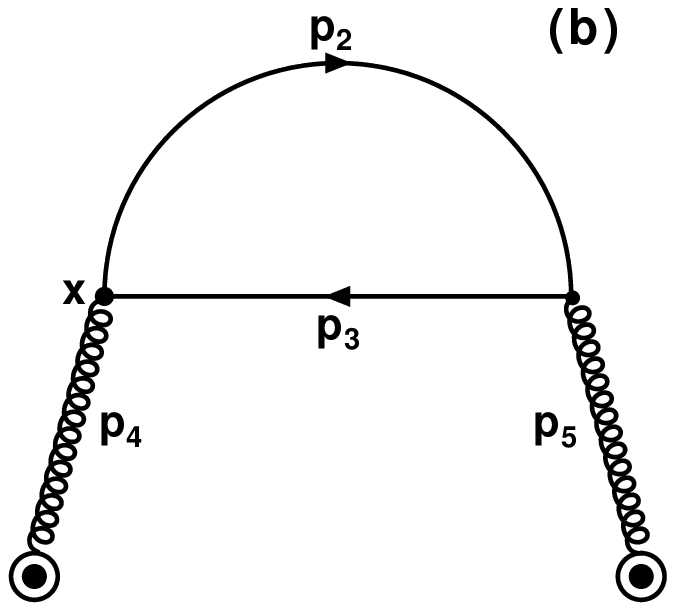}
}
\caption{Feynman diagrams contributing to $\langle \partial_\mu T^{\mu\nu}(x)\rangle$ in the presence of a source interaction term, $A_\mu^a \, j^\mu_a$.   The diagram in Figure 1(a) can be traced back to terms in the energy-momentum tensor (see equation (\ref{qemt})) which go as $\bar{\psi}\gamma\partial\psi$, whereas the diagram in Figure 1(b) originates from terms that go as $g\,\bar{\psi}\gamma A\psi$.  The contribution from the energy-momentum tensor and external current insertions are discussed in the text (see also Figure \ref{feyn2}).
}
\label{feyn1}
\end{figure*}

Performing a similar study within QCD currently seems out reach.  The problem is inherently dynamic and must be approached with real-time (and not imaginary-time) methods.  It is thus is not well suited for lattice techniques, which are more effective in imaginary time.  Even for a truly perturbative QGP, it is not clear how to sum all of the diagrams necessary to calculate the medium response in regions far from the fast parton.  In view of these limitations, one can consider two (of perhaps several) possibilities: a) calculate components of the EMT within perturbation theory to obtain information about the medium response in regions near the fast parton; b) calculate the source term for the EMT within perturbation theory and use an effective theory to propagate the resulting disturbance to regions far from the fast parton.  For this second option, one would calculate $\partial_\mu T^{\mu\nu} \equiv J^\nu$ using the fundamental EMT of QCD, and then allow $\partial_\mu T_H^{\mu\nu} = J^\nu$, where $T_H^{\mu\nu}$ is the hydrodynamic EMT, for instance, and then solve the resulting equations of motion.

Of the two options presented above, a) is under more theoretical control and will be considered in a future work.  In this work, however, I consider option b) because it is a more straightforward calculation and makes more direct contact with previous studies.  Furthermore, I will consider a medium of massless quarks/antiquarks, leaving the inclusion of medium gluons for a forthcoming study (thus, apart from the coupling strengths, the results will be the same as for a QED plasma, for which they are also new).  The EMT is given by \cite{qedemt}
\eqf{\label{qemt}
T^{\mu\nu} = \frac{i}{4}\bar{\psi}\left(\gamma^\mu\,\overset{\text{\tiny$\leftrightarrow$}}{D^\nu} + \gamma^\nu\,\overset{\text{\tiny$\leftrightarrow$}}{D^\mu}\right)\psi - g^{\mu\nu}\cur{L},
}
where
\eqf{
\cur{L} = \frac{i}{2}\bar{\psi}\,\overset{\text{\tiny$\leftrightarrow$}}{\slashed{D}}\,\psi \text{,     }D^\mu = \partial^\mu - i g\,A_a^\mu\,t^a
}
and
\eqf{
\bar{\psi}\,\gamma^\mu\,\overset{\text{\tiny$\leftrightarrow$}}{D^\nu} \, \psi = \bar{\psi}\,\gamma^\mu\,\overset{\text{\tiny$\rightarrow$}}{D^\nu} \, \psi - \bar{\psi}\,\gamma^\mu\,\overset{\text{\tiny$\leftarrow$}}{D^{*\nu}} \, \psi.
}
In the above equations, $g$ is the strong coupling, $t^a$ are the $SU(3)$ generators in the fundamental representation, and conventional slashed notation is used, $\slashed{A} = \gamma_\mu A^\mu$, etc.  A summation over color, spin, and the active number of quark flavors is implied in the EMT.

In order to investigate the medium response it is necessary to specify how the fast parton couples to the medium.  One possibility, which will be adopted here, is to model the fast parton as an external current which couples to the Lagrangian:
\eqf{\label{couplesource}
\cur{L} \rightarrow \cur{L} - A_\mu^a \, j^\mu_a,
}
where for the moment I do not specify the explicit form of $j$ ($j^\nu$ should not be confused with the source term, $J^\nu$).  The replacement made in (\ref{couplesource}) preserves non-Abelian gauge symmetry as long as $D^{ab}_\mu j^\mu_b = 0$.

The lowest order Feynman diagrams for calculating the thermal expectation $\langle \partial_\mu T^{\mu\nu}(x)\rangle$ in the presence of the interaction term $A_\mu^a \, j^\mu_a$, are shown in Figure \ref{feyn1}.  The two gluon exchange is necessary to couple to the EMT, which is a color singlet quantity (a two photon exchange is also necessary in QED, from Furry's theorem).   The diagram in Figure 1(a) arises from terms in (\ref{qemt}) which go as $\bar{\psi}\gamma\partial\psi$, whereas Figure 1(b) arises from terms that go as  $g\,\bar{\psi}\gamma A\psi$.  My convention is that the standard Feynman quark-gluon vertex contributes $i g \gamma^\mu t^a$, and I will use the Feynman gauge for gluon propagators.

\begin{figure}
\centerline{
\includegraphics[width = 0.35\linewidth]{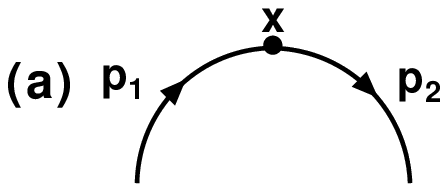}\hskip0.05\linewidth
\includegraphics[width = 0.26\linewidth]{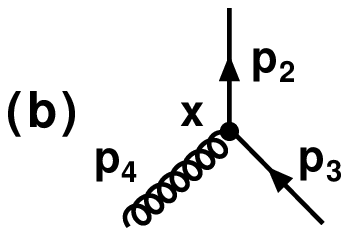}\hskip0.05\linewidth
\includegraphics[width = 0.17\linewidth]{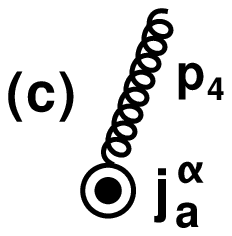}
}
\caption{Contributions to the Feynman diagrams of Figure \ref{feyn1} which come from inserting the energy-momentum tensor and external current interaction term.  Their values and how they are obtained are discussed in the text, see specifically equations (\ref{fig2a}) - (\ref{sourcerule}).}
\label{feyn2}
\end{figure}

In order to assign a value to these diagrams, one must determine what the correct Feynman rules for the EMT and external current are.  These nonstandard contributions are isolated in Figure \ref{feyn2}.  The contribution from the EMT shown in Figure  \ref{feyn2}(a) can be determined by simply assigning the appropriate momentum to each derivative.  Recalling that one is here interested in $\langle \partial_\mu T^{\mu\nu}\rangle$ and that the final result is Fourier transformed into position space, I find that the value of Figure \ref{feyn2}(a) is
\eqf{\label{fig2a}
&\frac{i e^{-i x \cdot(p_1 - p_2)}}{4} \\
&\times\left((p_2^2 - p_1^2) \gamma^\nu  + p_1^\nu (3 \slashed{p}_2 + \slashed{p}_1) -  p_2^\nu(3\slashed{p}_1 + \slashed{p}_2) \right).
}
One can apply the same procedure to Figure \ref{feyn2}(b) where I will choose the convention that the gluon momentum flows away from the external current (or into $x$ and any vertex) in all cases.  The result is
\eqf{\label{fig2b}
-&i g \,e^{-ix\cdot(p_4 + p_3 - p_2)}\,(p_4 + p_3 - p_2)_\mu \\
&\times\frac{\left(\gamma^\nu j_a^\mu + \gamma^\mu j_a^\nu - 2 g^{\mu\nu}\slashed{j}_a\right)t^a}{2}
}
which is only valid when the gluon field in Figure \ref{feyn2}(b) connects with the source.

Finally, for the source contribution in Figure \ref{feyn2}(c), one has very generally
\eqf{\label{sourcerule}
-i\int d^4 z \,j^\alpha_a(z) \, e^{i z\cdot p_4}
}
for the case of an external current which contains one power of $g$.  For the sake of simplicity and to be able to more easily compare with previous results, I will here consider an asymptotically propagating fast parton represented by $j^\mu_a = g Q^a(t) U^\mu\,\delta^3(\bald{z} - \bald{u}\,t)$ where $\bald{u}$ is the fast parton's velocity and $U^\mu = (1,\bald{u})$.  $g Q^a(t)$ is the charge of a classical particle in QCD (defined by $Q^a_i\,Q^a_j = \delta_{ij} C_{2i}$, with $C_{2i}$ the quadratic Casimir in representation $i$ (3 for a gluon, 4/3 for a quark)) and $Q^a(t)$ evolves in time according to Wong's equations \cite{Wong:1970fu}.  I will not consider the time dependence of $Q^a(t)$ here, as it is of higher order in $g$ (however, for higher order calculations one must account for the time dependence of $Q^a(t)$ in order to preserve gauge invariance).  For this choice of $j^\mu_a$ Figure \ref{feyn2}(b) simplifies to
\eqf{\label{easysourcerule}
-2\pi i \,g \, Q^a U^\alpha\,\delta(p_4\cdot U).
}
Before moving on, I point out once again that the Feynman gauge has been used in obtaining equations (\ref{fig2b}-\ref{sourcerule}).

It is worthwhile to mention here that the above results can be modified in a straightforward manner to consider more general situations.  For example, one could consider a fast parton created at an initial time, $t = 0$, by enforcing a step function in the external current or even include medium induced gluon radiation from the fast parton in a phenomenological way by treating the time dependent spectrum of energetic gluons as additional external currents in the medium.  In any of these situations, the rule expressed in (\ref{sourcerule}) can be applied.

As mentioned previously, it is necessary to here work in the real-time formalism of finite temperature field theory.  The Green's function structure of Figure \ref{feyn1} is obtained using the conventions of real-time thermal field theory as outlined by Das \cite{Das:1997gg}.  In the real-time formalism at finite temperature one must sum over two types of vertices which differ in value by an overall sign.  Additionally, propagators separate into vacuum and finite temperature contributions.  I will not present the details of the calculation in this brief report, reserving them instead for a follow-up publication in which medium gluons will also be considered.  Using standard Feynman rules for real-time thermal field theory in addition to the special rules presented in equations (\ref{fig2a}) - (\ref{sourcerule}) the result for the thermal contribution is obtained as
\begin{widetext}
\eqf{\label{fullsourceterm}
\langle \partial_\mu T^{\mu\nu}(x)\rangle &= -4 i \,N_F\,C_2\,g^4 \int \frac{d^4 p_3\,d^4 p_4\,d^4 p_5}{(2\pi)^{9}} e^{-i x \cdot(p_4 + p_5)} \delta(p_3^2)n_F(p_3)G_R(p_4)G_R(p_3 + p_4)G_R(p_5)\delta(p_4\cdot U)\delta(p_5\cdot U) \\
&\times\left[2\,(p_3\cdot U)^2\,p_5^\nu - U^2 \,p_3\cdot p_4\,p_5^\nu - U^\nu\,(p_3\cdot U)(2 p_3\cdot p_5 + p_4\cdot p_5)\right]
}
\end{widetext}
where $G_{R}(p) = (p^2 + i\epsilon p^0)^{-1}$ is the retarded Green's function, $n_F(p) = (e^{|p^0|/T} + 1)^{-1}$ is the Fermi distribution function, $T$ is the temperature, and $N_F$ is the number of active flavors.

Equation (\ref{fullsourceterm}) is the central result of this paper and it is worthwhile to consider the Green's function structure it contains, because this will provide an intuitive picture of the underlying physics.  If one reads from left to right in the first line it is clear that a particle from the heat bath (represented by the thermal distribution $ \delta(p_3^2)n_F(p_3)$) absorbs a gluon from the external current (represented by $G_R(p_4)$) and then continues to propagate (shown by $G_R(p_3 + p_4)$) until it absorbs another gluon from the external current (represented by $G_R(p_5)$).  A final propagator representing the reabsorption of the particle by the heat bath (which would be given by $G_R(p_3 + p_4 + p_5)$) has been eliminated by the momentum structure of $\langle \partial_\mu T^{\mu\nu}\rangle$.

In order to further extract meaningful information from equation (\ref{fullsourceterm}) it is useful to consider certain approximations and limits.   As was mentioned above, a kinetic theory calculation of the source term generated by an asymptotically propagating fast parton was performed in \cite{kineticsource}.  Equation (\ref{fullsourceterm}) should reduce to the kinetic theory result within the hard thermal loop (HTL) approximation (the HTL approximation is formally equivalent to the Vlasov equation \cite{Kelly:1994dh}).  It is worthwhile to check that this is indeed the case, as it gives confidence to the approach introduced in this paper and that (\ref{fullsourceterm}) has been evaluated correctly.

The HTL result can be obtained in the limit that the fields generated by the external current are soft compared to the temperature, or formally $|p_4| \sim g T \ll T$ in equation (\ref{fullsourceterm}) ($g$ is formally considered to be much less than 1).  Explicitly, one should expand $G_R(p_3 + p_4)$ in the limit of $|p_4| \ll |p_3|$ (since $|p_3|$, which appears in the Fermi distribution function, is cut off in the integration at a value on the order of $T$) to lowest contributing order.  With this guidance, I find that in the HTL approximation (\ref{fullsourceterm}) reduces to
\begin{widetext}
\eqf{\label{htlsourceterm}
\langle \partial_\mu T^{\mu\nu}(x)\rangle_{\text{HTL}} &= \frac{i \,m_D^2\,C_2\,g^2}{2} \int \frac{d^4 p_4\,d^4 p_5}{(2\pi)^{6}}\int\frac{d\Omega}{4 \pi} e^{-i x \cdot(p_4 + p_5)} G_R(p_4)G_R(p_5)\delta(p_4\cdot U)\delta(p_5\cdot U) \\
&\times\left[\frac{p_4^2\left((v\cdot U)^2\,p_5^\nu - U^\nu\,(v\cdot U)(v\cdot p_5)\right)}{(p_4^0 - \bald{v}\cdot\bald{p}_4 + i \epsilon )^2} + \frac{U^2 \,v\cdot p_4\,p_5^\nu + U^\nu\,(v\cdot U)(p_4\cdot p_5)}{p_4^0 - \bald{v}\cdot\bald{p}_4 + i \epsilon}\right]
}
\end{widetext}
where $m_D^2 = g^2\,T^2 \,N_F/6$ (quarks only).  In equation (\ref{htlsourceterm}) $d\Omega$ is the integration measure over the solid angle defined by the unit vector $\bald{v}$, and $v^\mu = (1,\bald{v})$.  Additionally, bare propagators are used for the Green's functions $G_R(p_4)G_R(p_5)$ in (\ref{htlsourceterm}) in order to facilitate comparison to previous results.

The interested reader can verify (as this author has done) that (\ref{htlsourceterm}) indeed reproduces the result of \cite{kineticsource}.  Rather than prove this for the full spatial distribution, I will here consider the total rate of energy transferred to the medium and show that this quantity matches with the previous result (leading log HTL).  The total rate of energy transfer, here denoted $dE/dt$, is found by integrating the zero component of the source term over all space
\eqf{\label{energyconservation}
\frac{d E}{d t} = \int d^3 x \, \langle \partial_\mu T^{\mu 0}(x)\rangle.
}
The spatial integration yields a factor $(2\pi)^3\delta(\bald{p}_4 + \bald{p}_5)$ from which one can trivially perform the $\bald{p}_5$ integration.  It is then straightforward to check that the two terms with coefficient $U^\nu$ cancel each other, and furthermore the term with coefficient $U^2$ will vanish by symmetry upon integration.  One finds the expression depends logarithmically on $|p_4|$.  The final result including the angular integrations is
\eqf{\label{dedthtl}
\frac{d E}{d t}_{\text{HTL}} &= \frac{m_D^2\,C_2\,\alpha_s}{2}\left(1 - \frac{\tanh^{-1} [u]}{\gamma^2 \,u}\right)\ln\frac{p_{max}}{p_{min}}
}
where $\gamma = (1-u^2)^{-1/2}$ and based on the approximation used in obtaining (\ref{htlsourceterm}) and physical reasoning $p_{max} \sim T$ and $p_{min}\sim m_D$.  The expression (\ref{dedthtl}) matches the leading log result of \cite{kineticsource}.  In what follows I will only consider the relativistic limit ($\gamma\gg1$) of (\ref{dedthtl}).

Having verified that (\ref{htlsourceterm}) reproduces known results from the HTL approximation, one can consider the full result, (\ref{fullsourceterm}), which includes contributions from the full momentum spectrum (that is, $|p_4| \sim |p_3|$ and $|p_4| \gg |p_3|$).  In order to simplify the discussion I will consider the source term ansatz discussed in \cite{neufrenk}
\eqf{\label{simplesource}
\langle \partial_\mu T^{\mu\nu}(x)\rangle \approx  \frac{d E}{d t}\left(U^\nu - \lambda \partial^\nu \right)\delta(\bald{x} - \bald{u} t).
}
The utility of (\ref{simplesource}) is that one can encode basic features of the source term in a compact way.  It is straightforward to check that (\ref{energyconservation}) is satisfied by (\ref{simplesource}).  The dimensionfull coefficient $\lambda$ parameterizes local contributions from the source, that is, terms which globally integrate to zero (and for instance do not contribute to $dE/dt$) but may still be important for exciting the medium.  It was found in \cite{neufrenk} that within linearized hydrodynamics a double peaked structure in the azimuthal emission spectrum associated with the source given by (\ref{simplesource}) only appeared for rather large values of $\lambda$ (on the order of 0.5 fm or higher for 20 GeV total energy deposited into the medium).  The basic assumption of equation (\ref{simplesource}) is that the source term (\ref{fullsourceterm}) can be expanded in terms of a $\delta$ function and derivatives of a delta function, which should be a reasonable approximation within the context of hydrodynamics.

It is straightforward to verify that the full source term given by equation (\ref{fullsourceterm}) can be put into the form of (\ref{simplesource}) by using  (\ref{energyconservation}) to obtain $dE/dt$ and
\eqf{\label{lamform}
\lambda \approx \frac{-1}{\frac{dE}{dt}}\int d^3 x\,x\,\langle \partial_\mu T^{\mu x}(x)\rangle
}
where I have chosen to use the $x$ component of the source to obtain $\lambda$ (according to the ansatz of (\ref{simplesource}) any component would work).  If one inserts  equation (\ref{lamform}) into (\ref{simplesource}) it is clear that $dE/dt$ drops out of the term proportional to the gradient, thus $dE/dt$ is simply a multiplicative factor in the evaluation of $\lambda$ (it is written in this way to make easier comparison to \cite{neufrenk}).  For this reason, and to simplify the analysis, I will take $dE/dt \approx dE/dt_{\text{HTL}}$ (I here remind the reader I use the relativistic limit of $dE/dt$ so that it contains no dependence upon the velocity of the fast parton) and instead focus on evaluating the integral term in (\ref{lamform}), for which I will use the full expression given by (\ref{fullsourceterm}).

The evaluation of the integral in (\ref{lamform}) can be done directly from (\ref{fullsourceterm}) through an integration by parts of the general form:
\eqf{\label{partsgrad}
\int d^3 x \,d^3 p\,x\, e^{i \bald{p} \cdot \bald{x}}\,f(\bald{p}) = i \int d^3 x \,d^3 p\,e^{i \bald{p} \cdot \bald{x}}\,\partial_{p_x} f(\bald{p}).
}
One has to be careful in using (\ref{partsgrad}) because $\lambda$ has an infrared divergence.  The simplest way around this, and what I will adopt here, is to introduce a mass term in the Green's functions $G_R(p_4)G_R(p_5)$ in (\ref{fullsourceterm})  given by the thermal gluon mass $m^2 = m_D^2/2$.

One can only apply the full result of equation (\ref{fullsourceterm}) down to some momentum scale $|p_4| \sim q^*$ where $g T \ll q^* \ll T$.  When $|p_4| \sim g T$ the full result contains contributions from higher order terms and only the HTL approximated result of equation (\ref{htlsourceterm}) is consistent to the order I am working.  From a practical point of view in the calculation of $\lambda$ this means dividing terms in the integration such that
\eqf{
\lambda &= \lambda_\text{F} \text{ when } |p_4|\geq q^*, \\
\lambda &= \lambda_{\text{HTL}} \text{ when } |p_4|\leq q^*,
}
where $ \lambda_\text{F}$ is obtained from (\ref{fullsourceterm}) and $\lambda_{\text{HTL}}$ is obtained from (\ref{htlsourceterm}).  The final result should be independent of $q^*$ (this type of analysis for the separation of scales in calculations of finite temperature field theory was introduced in \cite{Braaten:1991dd}).

\begin{figure}
\centerline{
\includegraphics[width = 0.75\linewidth]{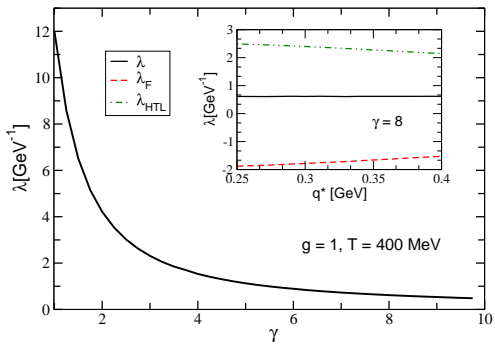}
}
\caption{In the source term of equation (\ref{simplesource}) $\lambda$ couples directly to sound modes and is important for generating observable Mach cone signals.  The contribution to $\lambda$ from the source term (\ref{fullsourceterm}) is plotted above for $g = 1$ and $T = 400$ MeV.  It is clear that $\lambda$ drops as a function of $\gamma = E/M$ ($M$ being the mass of the fast parton) which could have implications for the trigger $p_T$ dependence of azimuthal dihadron correlation measurements (see discussion in text).   The inset shows that the result for $\lambda$ is largely independent of the separation parameter, $q^*$ (see text for details).
}
\label{lambda}
\end{figure}

As mentioned above it is not possible to present the details of the calculation in this brief report and I will instead skip to the result for $\lambda$ and show that it is independent of $q^*$.  The result is shown in Figure \ref{lambda} as a function of $\gamma = (1-u^2)^{-1/2}$, and, at the risk of pushing the formal limit of $g\ll 1$, I have chosen the experimentally relevant values of $g = 1$ and $T = 400$ MeV.  It is interesting to note that $\lambda$ falls rather sharply as a function of $\gamma$.  Recall that $\lambda$ parameterizes local contributions from the source and is important for exciting the medium.  In particular, $\lambda$ couples directly to sound modes (and not to diffusive modes) when (\ref{simplesource}) is used as a source term for hydrodynamics and was found to be crucial to the appearance of a conical Mach-like emission spectrum in \cite{neufrenk}.  The experimental implication of the dependence of $\lambda$ on $\gamma$ as shown in Figure \ref{lambda} could be found in the trigger $p_T$ dependence of measurements of azimuthal dihadron particle correlations.  In particular, a conical emission pattern would be less likely to be observed for increasing trigger $p_T$, which indeed seems to be the case \cite{Adare:2010ry}.  The inset of Figure \ref{lambda} shows that the result is largely independent of the separation parameter, $q^*$, introduced above.

In summary I have calculated the source term for a medium of thermalized quarks in the presence of an asymptotically propagating fast parton directly from the EMT to leading order in perturbation theory and have presented the result in equation (\ref{fullsourceterm}).  Feynman rules specific to this calculation were presented in equations (\ref{fig2a}) - (\ref{sourcerule}) which can be applied in a straightforward way to more complicated external currents in future studies.  Within the HTL approximation, it was shown that (\ref{fullsourceterm}) reduces to a previous result \cite{kineticsource} derived from kinetic theory.  In addition to the HTL approximation, I considered the full result, which includes contributions when the source fields are hard (that is, $|p| \sim T$ and $|p| \gg T$).  In order to simplify the analysis, the source was put into the form of equation (\ref{simplesource}), in which case the source is described by the total rate of energy transferred to the medium, $dE/dt$, and local excitations described by the coefficient $\lambda$.  I here focused on the evaluation of $\lambda$ and found that it falls sharply as a function of $\gamma$, as shown in Figure \ref{lambda}.  This may have implications for the trigger $p_T$ dependence of measurements of azimuthal dihadron particle correlations in heavy-ion collisions.  In particular, a conical emission pattern would be less likely to be observed for increasing trigger $p_T$, which may indeed be the case \cite{Adare:2010ry}.  A future publication will include medium gluons and present the details of the calculation \cite{neuvitev}.

\small{{\it Acknowledgments}: I wish to thank Margaret Carrington, Berndt M\"uller, and Ivan Vitev for helpful comments and discussion.  This work was supported in part by the US Department of Energy, Office of Science, under Contract No. DE-AC52-06NA25396.}

\end{document}